\documentclass[prd,preprintnumbers,showkeys,nofootinbib]{article}
\usepackage{graphicx,amsmath,amssymb,epsf}

\begin{document}

\title{\Large \bf Exclusive central production of heavy quarks at the LHC}

\author{\large G. Chachamis$^1$,  M.~Hentschinski${}^{2,3}$, A. Sabio Vera${}^{2}$ , C. Salas${}^{2}$ 
\bigskip  \\
{\it  
${}^1$ Paul Scherrer Institut, CH-5232 Villigen PSI, Switzerland }
\bigskip \\
{\it  
${}^2$ Instituto de F{\'\i}sica Te\'orica UAM/CSIC,} \\
{\it Universidad Aut\'onoma de Madrid, Madrid, Spain}
\bigskip  \\
{ \it ${}^3$
 II. Institut f\"ur  Theoretische Physik,}\\
{\it  Universit\"at Hamburg,  Hamburg, Germany}\\
}

\maketitle

\vspace{-9cm}
\begin{flushright}
IFT-UAM/CSIC-09-56
\end{flushright}
\vspace{7.5cm}

\begin{abstract}
  We study the exclusive production of heavy flavors at central
  rapidities in hadron-hadron collisions within the $k_T$
  factorisation formalism. 
  Since this involves regions of small Bjorken $x$ in the 
  unintegrated gluon densities, we include the next-to-leading order BFKL 
  contributions working directly in transverse momentum representation. 
  Our results are presented in a form suitable for Monte Carlo 
  implementation. 

\end{abstract}

\section{Introduction}
\label{sec:intro}
Scattering processes with at least one hard scale are typically well 
described using perturbative QCD in the framework of collinear factorisation. 
In this approach cross sections are written as a convolution of a 
purely perturbative partonic cross section with non-perturbative parton 
distribution functions. The latter follow the DGLAP evolution
which describes their dependence on the hard perturbative scale. When 
the center of mass energy is very large compared to the perturbative hard 
scale, or a final state is fixed such that there are large rapidity 
differences among the emitted particles, an alternative high energy 
factorisation based on the Balitsky-Fadin-Kuraev-Lipatov (BFKL) evolution 
equation applies~\cite{bfkl}. Here the hard subprocess is convoluted 
with the hadron structure using unintegrated gluon densities which include  
their $k_T$ dependence in the small Bjorken $x$ limit. This can be seen 
as the small $x$ limit of $k_T$ factorisation. In this case enhanced 
logarithmic in $x$ contributions are resummed. 

In the present letter we propose to take the exclusive production of heavy 
quark-antiquark pairs in the early data at the LHC as a test of this 
formalism and the use of unintegrated gluon densities. Large masses such as 
those of bottom or top quarks allow for a perturbative treatment. In the 
case of top quark pairs their masses are so large that the typical proved 
values of Bjorken $x$ are not that small. In this case it is known that 
cross sections receive significant corrections from threshold 
logarithms (see~\cite{Bonciani:1998vc,Moch:2008qy,Cacciari:2008zb,Kidonakis:2008mu,Kidonakis:2009mx,Czakon:2008cx,Czakon:2009zw,Nason:1987xz,Beenakker:1990maa} for 
recent results in this direction). The bottom quarks are lighter and 
therefore test regions of smaller values of $x$ where the corresponding 
resummations find their natural environment. Previous investigations 
of heavy quark production similar to our present calculation where presented 
in~\cite{Ball:2001pq}. What we will show in this letter is an alternative 
approach which operates with NLO unintegrated gluon densities in transverse 
momentum space, does not involve the use of anomalous dimensions, treats the 
kinematics of the quark-antiquark pair exclusively and is 
readily suitable for a Monte Carlo analysis which we will present elsewhere.  
Other works which we found of interest in the field of inclusive heavy 
flavor production are Refs.~\cite{Ball:2001pq,Collins:1991ty,Catani:1990eg,Levin:1991ry,Baranov:2000gv,Baranov:2003cd,Jung:2001rp}. For the 
production of bottom pairs in particular we highlight 
Refs.~\cite{Shabelski:2004qy,Hagler:2000dda}, where the reported agreement
with experimental data at the Tevatron ranges from reasonable
\cite{Shabelski:2004qy} to very good \cite{Hagler:2000dda}.

The fully exclusive study that we propose could be very useful 
at the LHC since it allows for the precise determination of the $x$ values 
at which the unintegrated gluon densities are probed. This provides a 
good control on the accuracy of the approximations that we use in our 
calculation. At the LHC the dominant production process for both top and
bottom pairs is given by gluon-gluon fusion. However, as we already pointed 
out, only the bottom pair production occurs at relatively small $x$
providing the correct kinematics to apply high energy $k_T$ factorisation.  
Top pair production, on the other hand, occurs at relatively large values 
of $x$ due to the large top mass. Studies of its exclusive production
would certainly require the matching of the present calculation 
with renormalization group evolution and 
can be therefore considered as a test of the capability to extend our high 
energy factorisation towards the region of large $x$. For our predictions we 
incorporate the NLO corrections to the BFKL evolution 
kernel~\cite{nlla}. A related study, devoted to the exclusive central 
production of jets in hadron-hadron collisions in $k_T$ factorisation, 
was presented in Ref.~\cite{Bartels:2006hg}.

After this brief Introduction, in Section 2, we present the general 
structure of the $k_T$ factorised differential cross-section and calculate  
its different 
elements. In Section 3 we discuss the unintegrated gluon density in 
$k_T$ space and its iterative structure. Finally, we write our Conclusions 
in Section 4.

\section{The $k_T$ factorised differential cross-section at NLO}
\label{sec:results}
To describe the differential cross-section for the exclusive production of
a pair of heavy quarks within $k_T$ factorisation it is convenient to
introduce a Sudakov basis. To this end we define the light-like momenta
$p_1$ and $p_2$ which coincide in the $s \to \infty$ limit with the
momenta of the incoming protons $p_{A}$ and $p_{B}$:
\begin{equation}
  \label{eq:p12}
p_1 ~=~ p_A - \frac{m_P^2}{s} p_B, \hspace{1cm} 
p_2 ~=~ p_B - \frac{m_P^2}{s} p_A,
\end{equation}
with $s = (p_A + p_B)^2$ being the squared center of mass energy of the
hadronic process. With these definitions, we can then work with the
usual Sudakov decomposition of a general four momentum, {\it i.e.}
\begin{align}
  \label{eq:sudakov}
k = \alpha \, p_1 + \beta \, p_2 + k_{\perp}.
\end{align}

\begin{figure}[htbp]
  \centering
  \parbox{3cm}{\includegraphics[width=5cm]{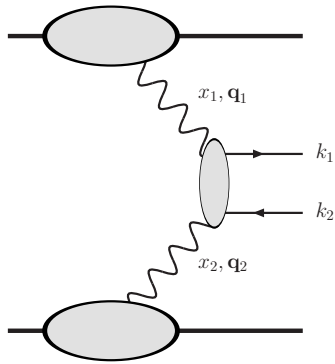}}
  \caption{\small Central production of 
two heavy quarks in $k_T$ factorisation}.
  \label{fig:process}
\end{figure}

The notation for the relevant momenta in the partonic hard subprocess 
is given in Fig.~\ref{fig:process}. In the BFKL formalism $t$-channel 
gluons carry a modified propagator which reggeises them. This propagator 
is associated to the momenta $q_1$ and $q_2$ in Fig.~\ref{fig:process}. 
These simplify in the high energy limit and can be written as
\begin{equation}
  \label{eq:reggeised_gluon}
q_1 ~=~ x_1 \, p_1 + q_{1,\perp}, \hspace{1cm} 
q_2 ~=~ x_2 \, p_2 + q_{2,\perp}.
\end{equation}
On the other hand, the momenta of the produced heavy quarks have the
following decomposition
\begin{equation}
  \label{eq:producedquarks}
  k_{i} ~=~ \alpha_{i} \, p_1 + \beta_i \, p_2 + k_{i,\perp},
\hspace{1cm} i ~=~ 1,2.
\end{equation}
Taking into account the on-shellness of the produced quarks, the above
Sudakov parameters can be expressed in terms of rapidities, transverse 
momenta and heavy quark masses $M$, {\it i.e.}
\begin{equation}
  \label{eq:alpha}
\alpha_{i} ~=~  \sqrt{\frac{M^2 + {\bf k}_{i}^2 }{s}} e^{\eta_{i}},
\hspace{1cm}
\beta_{i} ~=~  \sqrt{\frac{M^2 + {\bf k}_{i}^2 }{s}} e^{-\eta_{i}},
\hspace{1cm} i = 1,2.
\end{equation}
Here $\eta_1$ ($\eta_2$) is the rapidity of the produced heavy quark 
(anti-quark) and ${\bf k}_i^2 = - k_{i, \perp}^2$ are the corresponding 
Euclidean squared transverse momenta. 

Making use of the definitions 
\begin{equation}
  \label{eq:sab}
s_1 ~=~ (p_1 + q_2)^2 = x_2 \, s, \hspace{1cm}
s_2 ~=~ (p_2 + q_1)^2 = x_1 \, s,
 \end{equation}
 which correspond to the center of mass energies of the upper and
 lower subamplitudes in Fig.~\ref{fig:process}, respectively, we can write the
 following expression for the differential cross-section of heavy quark
 production:
\begin{align}
  \label{eq:reggefac}
\frac{d^6\sigma}{d\eta_1 d\eta_2 d^2{\bf k}_1 d^2{\bf k}_2} = &
 \int_0^1 d x_1 \int_0^1 d x_2  \int \frac{d^2 {\bf q}_1}{(2\pi)^3} 
 \int \frac{d^2 {\bf q}_2}{(2\pi)^3} 
 \left[ \int \frac{d^2 {\bf q}_a}{2\pi} 
 \frac{\Phi_A({\bf q}_a)}{{\bf q}_a^2} 
 f\left(\frac{s_1}{s_{0,1}}, {\bf q}_a, {\bf q}_1\right) 
\right]
\notag \\
\times&
\frac{|\Gamma_{\text{RR}\to \text{Q}\bar{\text{Q}}}({\bf q}_1, {\bf q}_2; {\bf k_1}, {\bf k}_2, z)|^2}{{\bf q}_1^2 {\bf q}_2^2}
\left[ \int \frac{d^2 {\bf q}_b}{2 \pi} 
 \frac{ \Phi_B({\bf q}_b)}{{\bf q}_b^2}  
f\left(\frac{s_2}{s_{0,2}}, {\bf q}_2, {\bf q}_b\right)
\right]
 \notag \\
\times&
(2\pi)^4\delta^{(2)}({\bf q}_1 + {\bf q}_2 - {\bf k}_1 - {\bf k_2}  ) 
 \, \delta(x_1 - \alpha_1 - \alpha_2) \, \delta(x_2 - \beta_1 - \beta_2). 
\end{align}
In this expression $\Phi_{A}$ and $\Phi_{B}$ denote the hadron 
impact factors, which are 
responsible for the coupling of the reggeised gluon to
the proton A and B, respectively. $\Gamma_{\text{RR}\to
  \text{Q}\bar{\text{Q}}}$ indicates the high energy effective vertex 
coupling the two reggeised gluons to the heavy quark-antiquark pair with
\begin{align}
  \label{eq:z}
z = \frac{\alpha_1}{x_1} = \frac{\sqrt{{\bf k}_1^2 + M^2}}{\sqrt{{\bf k}_1^2 + M^2} + \sqrt{{\bf k}_2^2 + M^2}e^{\eta_2 - \eta_1}}
\end{align}
being the fraction of the longitudinal momentum of the upper reggeised gluon
along $p_1$, carried by the heavy quark.  $f$ denotes the BFKL gluon Green 
function with the following Mellin transform $f_\omega$:
\begin{align}
  \label{eq:greensfunction}
 f\left(\frac{s_1}{s_{0,1}}, {\bf q}_a, {\bf q}_1\right) = \int_{\cal C} 
\frac{ d\omega}{2\pi i} \left(\frac{s_1}{s_{0,1}}\right)^{\omega}  f_\omega ({\bf q}_a, {\bf q}_1 ),
\end{align}
where the contour of integration ${\cal C}$ 
lies parallel to the imaginary axis and to
the right of all the singularities in $f_\omega$. The resummation of high 
energy logarithms is achieved by iterating the BFKL integral equation for 
$f_\omega$:
\begin{align}
  \label{eq:bfkl}
\omega  f_\omega ({\bf q}_a, {\bf q}_1 ) 
&=
\delta^{(2)} ({\bf q}_a - {\bf q}_1 ) + \int d^2 {\bf q} \, 
K_{\text{BFKL}} ({\bf q}_a, {\bf q})  \, 
f_\omega ({\bf q}, {\bf q}_1 ).
\end{align}
In a general case where both the produced heavy quarks and the impact
factors would provide a similar hard scale 
({\it i.e.} if the protons were replaced by highly virtual photons, or jets 
with a high $p_t$ were tagged in the forward/backward regions), a good 
choice for the energy scales $s_{0,i}$ would be given by 
$s_{0,1} = |{\bf q}_a|\sqrt{\Sigma}$ and 
$s_{0,2} = \sqrt{\Sigma}|{\bf q}_b|$, where
\begin{equation}
  \label{eq:Sigma}
\Sigma ~=~ x_1 \, x_2 \, s ~=~ \hat s + ({\bf k}_1 + {\bf k}_2)^2, 
\hspace{1cm}
\hat s ~=~ (k_1 + k_2)^2.
\end{equation}
$\hat s$ reads for the squared center of mass energy of the partonic
process $g^*g^* \to Q\bar{Q}$. Such a choice naturally introduces the
rapidities $\eta_{\tilde A} $ and $\eta_{\tilde B} $ of the emitted
particles with momenta $p_{\tilde A}$ and $p_{\tilde B}$ since
\begin{equation}
  \label{eq:rapidity}
\left( \frac{s_1}{s_{0,1}} \right)^{\omega} ~=~ e^{(\eta_{\tilde A} - \eta_{Q\bar{Q}})\omega}, \hspace{1cm}
\left( \frac{s_2}{s_{0,2}} \right)^{\omega} ~=~ e^{( \eta_{Q\bar{Q}} - \eta_{\tilde B})\omega},
\end{equation}
with the rapidity of the heavy quark system being given by
\begin{align}
  \label{eq:rap_heavyquark}
  \eta_{Q\bar{Q}} = \frac{1}{2} \ln \frac{\alpha_1 + \alpha_2}{\beta_1 + \beta_2}.
\end{align}
If the BFKL Green function, the impact factors and the
production vertex were known exactly at NLO, then the precise choice of the
energy scales $s_{0,i}$ would turn out to be irrelevant since 
any dependence of the cross section 
on this scale would cancel at the same NLO accuracy. However, 
even in that case, when the kernel is exponentiated there is a residual 
dependence on $s_{0,i}$ which would correspond to NNLO and higher terms. 
A natural choice for $s_{0,i}$ is then that which reduces the size of those 
higher orders corrections to the minimum for a given observable. 

In the case of interest for us in this letter there exists a hierarchy of 
scales with a large difference between the only hard scale 
provided by the invariant mass of the heavy quark pair system and the 
large transverse size of the incoming hadrons. Here the previous symmetric 
choice of scales is not appropriate as the scale of the heavy quark 
anti-quark system $\Sigma$ is significantly larger than the transverse 
scales ${\bf q}_a^2 $ and ${\bf q}_b^2$ associated to the scattered 
protons. A more natural choice for $s_{0,i}$ is given by $\Sigma$ alone, 
{\it i.e.}
\begin{equation}
  \label{eq:Bjorken}
\left( \frac{s_1}{s_{0,1}} \right)^{\omega} ~=~ x_1^{-\omega}, \hspace{1cm}
\left( \frac{s_2}{s_{0,2}} \right)^{\omega} ~=~ x_2^{-\omega}.
\end{equation}
This choice of the energy scale is common in deep inelastic scattering 
and leads to the concept of the unintegrated gluon density in a hadron. This 
represents the probability of resolving an off-shell gluon carrying a 
longitudinal momentum fraction $x$ off the incoming hadron, together 
with a transverse momentum $k_T$. 

As it is well-known, any choice of energy scale only matters at 
next-to-leading and 
higher orders since the LO approach is scale invariant. The 
LO unintegrated gluon density $g^\text{LO}$ is defined as
\begin{align}
  \label{eq:LOgluondensity}
g^{\text{LO}}(x, {\bf k}) &= \int  \frac{d^2 {\bf q}}{2\pi} \frac{\Phi_P({\bf q})}{{\bf q}^2} f^{\text{LO}}(x,{\bf q}, {\bf k} ),
& \text{with} &&
f^{\text{LO}}(x,{\bf q}, {\bf k} ) & =  \int_{\cal C} 
\frac{d\omega}{2\pi i} x^{-\omega}
f^{\text{LO}}_\omega ({\bf q}, {\bf k}),
\end{align}
where $f^{\text{LO}}_\omega$ corresponds to the solution of the LO BFKL 
equation with kernel $K_{\text{BFKL}} = K_{\text{BFKL}}^{\text{LO}} $.
Contrary to the LO case, the next-to-leading order BFKL evolution is 
sensitive to changes in the energy scales $s_{0, i}$. As it was pointed out 
in~\cite{Bartels:2006hg}, any shift of scales can be absorbed in the kernel, 
impact factors, and central production vertex. With the choice of energy 
scale as in Eq.~(\ref{eq:Bjorken}) the NLO impact factors are modified by an 
extra logarithmic term of the form
\begin{align}
\label{eq:nloimpa_mod}
  \tilde\Phi^{\text{NLO}}_P({\bf q}) = \Phi^{\text{NLO}}_P ({\bf q})
- \frac{{\bf q}^2}{2} \int d^2{\bf l} \, 
\frac{ \Phi^{\text{LO}}_P({\bf l}) }{ {\bf l}^2} \, 
K^{\text{LO}}_{\text{BFKL}} ({\bf l}, {\bf q}) \, 
\ln \frac{{\bf l}^2}{{\bf q}^2}.
\end{align}
The NLO kernel receives two additional contributions, corresponding to the 
incoming and outgoing reggeised gluons:
\begin{align}
  \label{eq:nlobfkl_mod}
\tilde K^{\text{NLO}}_{\text{BFKL}}({\bf l}_a , {\bf l}_b) =&
K^{\text{NLO}}_{\text{BFKL}}({\bf l}_a, {\bf l}_b) 
-\frac{1}{2} \int d^2{\bf l} \, 
 K^{\text{LO}}_{\text{BFKL}}({\bf l}_a, {\bf l}) \,
 K^{\text{LO}}_{\text{BFKL}}({\bf l}, {\bf l}_b) \,
\ln \frac{{\bf l}^2}{{\bf l}_b^2}.
\end{align}
The NLO $Q\bar{Q}$ production vertex also gets two types of 
corrections, corresponding to the two different evolution
chains originating from the hadrons A and B:
\begin{align}
  \label{eq:nlogamma_mod}
  |\tilde\Gamma_{ \text{RR} \to \text{Q} \bar {\text{Q}}}^{\text{NLO}}
 ({\bf q}_1, {\bf q}_2; &{\bf k}_1, {\bf k}_2, z)  |^2 
~=~ |\Gamma_{\text{RR} \to
    \text{Q} \bar{\text{Q}} }^{ \text{NLO} } ({\bf q}_1, {\bf q}_2; {\bf k}_1, {\bf k}_2, z)  |^2
\notag \\
&-
  \frac{ {\bf q}_1^2}{2} \int \frac{d^2 {\bf l}}{{\bf l}^2} \, 
  K_{\text{BFKL}}^{\text{LO}} ({\bf q}_1, {\bf l} ) \, 
  |\Gamma_{\text{RR} \to \text{Q} \bar{\text{Q}} }^{ \text{LO}}
    ({\bf l}, {\bf q}_2; {\bf k}_1, {\bf k}_2, z)|^2
  \, \ln{\frac{{\bf l}^2}{({\bf q}_2 + {\bf l})^2}}
\notag \\
&
- \frac{ {\bf q}_2^2}{2} \int \frac{d^2 {\bf l}}{{\bf l}^2}
   \, {|\Gamma_{\text{RR} \to \text{Q} \bar{\text{Q}} }^{ \text{LO}
    }({\bf q}_1, {\bf l}; {\bf k}_1, {\bf k}_2, z)|^2 } \,
 K_{\text{BFKL}}^{\text{LO}} ({\bf l}, {\bf q}_2 ) \,
 \ln{\frac{{\bf l}^2}{( {\bf q}_1 + {\bf l} )^2}}.
\end{align}
With these modifications, the NLO unintegrated gluon density is
defined as follows
 \begin{equation}
   \label{eq:NLOgluondensity}
 g^{\text{NLO}}(x, {\bf k}) 
~=~
 \int  \frac{d^2 {\bf q}}{2\pi} \frac{\tilde{\Phi}_P({\bf q})}{{\bf q}^2}  \tilde{f} (x, {\bf q}, {\bf k}), \hspace{1cm}
\tilde{f} (x, {\bf q}, {\bf k})
~=~
\int_{\cal C} \frac{d\omega}{2\pi i} x^{-\omega}
 \tilde{f}_\omega ({\bf q}, {\bf k}),
 \end{equation}
 where $\tilde{f}_\omega$ obeys the modified NLO BFKL equation
 \begin{align}
   \label{eq:bfkl_mod_nlo}
 \omega  \tilde{f}_\omega ({\bf q}_a, {\bf q}_1 ) 
 &=
 \delta^{(2)} ({\bf q}_a - {\bf q}_1 ) + \int d^2 {\bf q} \tilde{K}_{\text{BFKL}} ({\bf q}_a, {\bf q} ) 
 \tilde{f}_\omega ({\bf q}, {\bf q}_1 )
 \end{align}
 with a NLO kernel which we will discuss in Section 3:
 \begin{align}
   \label{eq:bfkllo+nlo}
  \tilde{K}_{\text{BFKL}} ({\bf q}_a, {\bf q} )  = {K}^{\text{LO}}_{\text{BFKL}} ({\bf q}_a, {\bf q} )   + \tilde{K}^{\text{NLO}}_{\text{BFKL}} ({\bf q}_a, {\bf q} ).
 \end{align}
Using these definitions, the differential cross section in 
Eq.~(\ref{eq:reggefac}) at NLO accuracy is given by the expression 
 \begin{align}
   \label{eq:reggefac2}
 & \frac{d^6\sigma}{d\eta_1 d\eta_2 d^2{\bf k}_1 d^2{\bf k}_2} = 
  \int_0^1 d x_1 \int_0^1 d x_2 \int \frac{d^2 {\bf q}_1}{(2\pi)^3} 
  \int \frac{d^2 {\bf q}_2}{(2\pi)^3}  \, 
g^{\text{NLO}}(x_1, {\bf q}_1)  \, 
g^{\text{NLO}}(x_2, {\bf q}_2)
 \notag \\
 & \qquad  \times 
\frac{|\Gamma_{\text{RR}\to \text{Q}\bar{\text{Q}}} ({\bf q}_1, {\bf q}_2; 
{\bf k}_1, {\bf k}_2, z)|^2}{{\bf q}_1^2 {\bf q}_2^2}  
\, \delta^{(2)}({\bf q}_1 + {\bf q}_2 - {\bf k}_1 - {\bf k}_2  ) 
\,  \delta(x_1 - \alpha_1 - \alpha_2) \, \delta(x_2 - \beta_1 - \beta_2) 
 \\
 &   = \int \frac{d^2 {\bf q}_1}{(2\pi)^6} 
  g^{\text{NLO}}(\alpha_1 + \alpha_2, {\bf q}_1)  
g^{\text{NLO}}(\beta_1 + \beta_2,  {\bf k}_1 + {\bf k_2} - {\bf q}_1)
  \frac{|\Gamma_{\text{RR}\to \text{Q}\bar{\text{Q}}} 
({\bf q}_1,  {\bf k}_1 + {\bf k}_2 - {\bf q}_1 ; 
{\bf k}_1, {\bf k}_2, z)|^2}{{\bf q}_1^2 ({\bf k}_1 + {\bf k_2} - 
{\bf q}_1)^2}. \notag
 \end{align}
This expression can be interpreted as the
 convolution of the unintegrated gluon densities with a partonic
 differential cross section {\it i.e.}
 \begin{align}
   \label{eq:partonic}
 \frac{d^6\sigma}{d\eta_1 d\eta_2 d^2{\bf k}_1 d^2{\bf k}_2}
 &=
 \int_0^1 \frac{d x_1}{x_1} \int_0^1  \frac{ d x_2}{x_2}   \int 
\frac{{d^2 {\bf q}_1}}{2\pi^2}
  \int \frac{{d^2 {\bf q}_2}}{2 \pi^2} \,  g^{\text{NLO}}(x_1, {\bf q}_1) \,  
g^{\text{NLO}}(x_2, {\bf q}_2) \, 
\frac{d^6\hat{\sigma}}{d\eta_1 d\eta_2 d^2{\bf k}_1 d^2{\bf k}_2}
 \end{align}
 where
 \begin{align}
   \label{eq:partonic_crossection}
  {d^6\hat{\sigma}} \equiv 
 \frac{1}{2\Sigma} |{\cal A}(q_1, q_2; k_1, k_2)|^2  
\frac{d\eta_1 d^2{\bf k}_1}{2 (2\pi)^3} 
\frac{d\eta_2 d^2{\bf k}_2}{2 (2\pi)^3}
(2\pi)^4 \delta^{(4)}(q_1 + q_2 -  k_1 - k_2 ).
 \end{align}
 Here  $2\Sigma$ is  the  flux factor and
 \begin{align}
   \label{eq:asquared}
 |{\cal A}(q_1, q_2; k_1, k_2)|^2 = \frac{\Sigma^2}{ {\bf q}_1^2 {\bf q}_2^2}{|\Gamma_{\text{RR}\to \text{Q}\bar{\text{Q}}} ({\bf q}_1, {\bf q}_2; {\bf k}_1, {\bf k}_2, z)|^2}
 \end{align}
 is the squared matrix element for the production of a heavy $Q\bar{Q}$ 
pair from the fusion of two {\it transversely} polarised
 reggeised gluons. This means that their polarisations are chosen to satisfy
 \begin{equation}
   \label{eq:polarization_sum_transverse}
   \sum_{\lambda}
   \epsilon_{(\lambda)}^{\mu}(q_i)\epsilon_{(\lambda)}^{\nu}(q_i) ~=~
   \frac{{\bf q}_i^\mu {\bf q}_i^\nu}{{\bf q}_i^2},  \hspace{1cm} 
 \text{with $i=1,2$},
 \end{equation}
 and can be related up to the overall factor $\Sigma^2/{\bf q}_1^2
 {\bf q}_2^2$ in Eq.~(\ref{eq:asquared}) to the usual longitudinally 
 polarised reggeised gluons by means of a Ward-identity for the
 $t$-channel gluons.
 
 At present the NLO corrections to the heavy quark production
 vertex $\Gamma_{\text{RR} \to \text{Q} \bar{\text{Q}}}$ are not
 available and only the LO vertex is known~\cite{Ball:2001pq}. 
This LO result can be written in the following form
\begin{align}
  \label{eq:gamma_squared}
|\Gamma^{\text{LO}}_{\text{RR} \to \text{Q} \bar{\text{Q}}}  ( {\bf q}_1, {\bf q}_2 ; {\bf k}_1, {\bf k}_2, z)|^2 = {g^4 } \left( \frac{N_c}{2} A_1( {\bf q}_1, {\bf q}_2 ; {\bf k}_1, {\bf k}_2, z) + \frac{1}{ 2N_c}  A_2( {\bf q}_1, {\bf q}_2 ; {\bf k}_1, {\bf k}_2, z) \right),
\end{align}
with
\begin{align}
  \label{eq:A1}
A_1  ( {\bf q}_1, {\bf q}_2 ; {\bf k}_1, {\bf k}_2, z) =& 
\frac{{\bf q}_1^2 {\bf q}_2^2}{\hat s \, \Sigma} 2  
\bigg(\frac{1}{\hat t - M^2} - \frac{1}{\hat u - M^2} \bigg)
\bigg(\frac{(1-z){\bf k}_1^2 + M^2 }{z} - \frac{z {\bf k}_2^2 + M^2}{1 - z} \bigg)
\notag \\
&- \bigg( 
      \frac{[({\bf q}_1 - {\bf k}_1)^2 + M^2][({\bf q}_1 - {\bf k}_2)^2 + M^2] - ({\bf k}_1^2 + M^2)({\bf k}_2^2 + M^2)}{(\hat t - M^2)(\hat u - M^2)}
\bigg)^2 
\notag \\
&+
\bigg(
\frac{({\bf q}_1 - {\bf k}_2)^2 + M^2 - \frac{z}{1 - z}({\bf k}_2^2 + M^2) }{\hat u - M^2} + \frac{E(M^2)}{\hat s}
\bigg)
\notag \\ & \qquad \qquad \qquad \qquad  \times
\bigg(
\frac{({\bf q}_1 - {\bf k}_1)^2 + M^2 - \frac{1- z}{ z}({\bf k}_1^2 + M^2) }{\hat t - M^2} - \frac{E(M^2)}{\hat s}
\bigg),
\end{align}
and
\begin{align}
  \label{eq:a2}
A_2( {\bf q}_1, {\bf q}_2 ; {\bf k}_1, {\bf k}_2, z) = & 
\bigg( 
      \frac{[({\bf q}_1 - {\bf k}_1)^2 + M^2][({\bf q}_1 - {\bf k}_2)^2 + M^2] - ({\bf k}_1^2 + M^2)({\bf k}_2^2 + M^2)}{(\hat t - M^2)(\hat u - M^2)}
\bigg)^2 
\notag \\
& 
-
\frac{{\bf q}_1^2 {\bf q}_2^2}{(\hat t - M^2)(\hat u - M^2)} .
\end{align}
To write down Eqs.~(\ref{eq:A1},\ref{eq:a2}) we defined, apart
from the variables introduced in Eqs.~(\ref{eq:z},\ref{eq:Sigma})
and the partonic Mandelstam invariants
\begin{align}
  \label{eq:mandelstam}
\hat t &= (q_1 - k_1 )^2 = -\frac{1-z}{z}({\bf k}_1^2 + M^2) - ({\bf q}_1 - {\bf k}_1)^2, \\
\hat u & = (q_1 - k_2)^2 = \frac{z}{1 -z} ({\bf k}_2^2 + M^2) - ({\bf q}_1 - {\bf k}_2)^2,
\end{align}
the following set of transverse momenta 
\begin{equation}
  \label{eq:conv2}
{\bf \Delta} ~=~ {\bf k}_1 + {\bf k}_2, \hspace{1cm}
{\bf \Lambda} ~=~ (1 - z) {\bf k}_1 - z{\bf k}_2,
 \end{equation}
which allow us to express Eq.~(\ref{eq:Sigma}) as
\begin{align}
\label{eq:Sigma_mod}
  \Sigma & = \hat{s}  + {\bf \Delta}^2 = \frac{{\bf \Lambda}^2 + M^2}{z(1- z)} + {\bf \Delta}^2.
\end{align}
Finally, we have also used
\begin{align}
  \label{eq:ee}
E(M^2) &\equiv 2(2z -1) {\bf q}_1^2 + 2 {\bf q}_1 \cdot {\bf \Lambda} 
        + \frac{1 - 2z}{z(1-z)} ({\bf \Lambda}^2 + M^2) - 
\left[(2z -1) {\bf \Delta}^2 + 2{\bf \Lambda}\cdot{\bf \Delta} \right] \frac{{\bf q}_1^2}{\Sigma}.
\end{align}

The explicit form of the vertex in Eq.~(\ref{eq:gamma_squared}), 
keeping all the information on the 
outgoing $Q\bar{Q}$ system, will permit a comprehensive study of 
differential distributions in exclusive observables. For this we will 
also need to keep track of the multiple soft emission stemming from the 
gluon evolution. How to achieve this task is discussed in the following 
Section. 

\section{Multiparticle production and the unintegrated gluon density}
\label{sec:ite}

The NLO unintegrated gluon densities in Eq.~(\ref{eq:NLOgluondensity}), 
which enter the differential cross section of Eq.~(\ref{eq:reggefac2}), 
require both the NLO BFKL gluon Green function and the proton impact factor
$\Phi_P({\bf q})$. The latter is of non-perturbative origin, it can
only be modelled and has to be extracted from the data. A possible simple 
choice for a model of the proton impact factor would be 
\begin{align}
  \label{eq:proton_model}
\Phi_P({\bf q}) &\sim 
\left( \frac{{\bf q}^2 }{{\bf q}^2 + \Lambda^2 } \right)^{\lambda}.
\end{align}
Here $\lambda$ is a positive free parameter, while $\Lambda$ is a momentum
scale of the order of $\Lambda_\text{QCD}$. A more sophisticated alternative to
Eq.~(\ref{eq:proton_model}) has been presented in Ref.~\cite{Ellis:2008yp}
where it was proposed to expand the proton impact factor over a set
of orthogonal conformal invariant eigenfunctions.

The second building block of the unintegrated gluon densities is given
by the NLO BFKL gluon Green function. An alternative formulation to the 
usual treatment in Mellin space was proposed in Ref.~\cite{Andersen:2003wy}. 
This form of solving the equation by iteration in momentum space has the 
advantage of dealing exactly with running coupling effects and incorporates 
the full azimuthal angle dependence of the soft multiparticle emission 
in multi-regge kinematics associated to the BFKL evolution. For a complete 
analysis of exclusive properties of multigluon final states associated to 
the production of a heavy $Q\bar{Q}$ pair we consider this to be the most 
convenient of the available methods of analysis of the BFKL Green function 
at NLO. By means of a phase space slicing parameter $\lambda$ the virtual 
and real contributions are treated separately. 

 As it has been previously explained, the BFKL kernel receives in the case 
 of NLO unintegrated gluon densities an additional contribution,
 Eq.~(\ref{eq:nlobfkl_mod}), due to the choice of the energy scales
 $s_{0,i}$. This affects the real emission contribution to the kernel 
but not the gluon Regge trajectory, $\omega_\lambda$, which in this 
physical regularisation can be written as  
\begin{equation}
   \label{eq:omega_lambda}
\omega_\lambda ({\bf q}) =
 -\xi (|{\bf q}|\lambda)\ln \frac{{\bf q}^2}{\lambda^2} + 
\bar{\alpha}_s^2 \frac{3}{2} \zeta(3)
\hspace{.5cm}
\text{with} \hspace{0.5cm}
\xi(X) = \bar{\alpha}_s + \frac{\bar{\alpha}_s^2}{4} 
\left( \frac{4}{3} - \frac{\pi^2}{3} + \frac{5}{3} \frac{\beta_0}{N_c} 
- \frac{\beta_0}{N_c} \ln \frac{X}{\mu^2}   \right), 
\end{equation}
where $\bar{\alpha}_s = \alpha_s(\mu) N_c / \pi$ and $\beta_0 = (11 N_c -
2 n_f)/3$.  $\mu$ is the renormalisation scale in the $\overline{\text{MS}}$
scheme. $\lambda$ can be understood as an effective gluon mass or as a 
lower cut-off for the transverse momenta of the emitted gluons. 

The NLO real emission kernel $\tilde{K}^{\text{real}}_{ \lambda} $, 
which accounts 
for the emission of gluons or massless quarks in quasi-multi-regge kinematics,
is given by the following sum
 \begin{align}
   \label{eq:kernel_lamgda}
 \tilde{K}^{\text{real}}_{ \lambda} ({\bf l}_a, {\bf l}_a +  {\bf l} )  = 
\frac{1}{\pi {\bf l}^2} \xi({\bf l}^2) \theta({\bf l}^2 - \lambda^2)
+ 
\hat{K}^{\text{real}} ({\bf l}_a, {\bf l}_a + {\bf l}) 
+
K_{\text{coll}} ({\bf l}_a, {\bf l}_a + {\bf l}),
 \end{align}
where~\cite{nlla}
\begin{align}
  \label{eq:real_finite}
   \hat{K}^{\text{real}} ({\bf l}_a, {\bf l}_b) 
=&
\frac{\bar{\alpha}_s^2}{4\pi} 
\bigg\{
- \frac{1}{({\bf l}_a - {\bf l}_b )^2} \ln^2 \frac{{\bf l}_a^2}{{\bf l}_b^2}
+
\left( 1 + \frac{n_f}{N_c^3}  \right)
 \left( \frac{3 ({\bf l}_a \cdot {\bf l}_b)^2 - 2 {\bf l}_a^2 {\bf l}_b^2}{16 {\bf l}_a^2 {\bf l}_b^2 }  \right)
\notag \\
&
\times
\left[ \frac{2}{ {\bf l}_a^2 }+  \frac{2}{ {\bf l}_b^2 } 
+
\left( \frac{1}{ {\bf l}_b^2 }-  \frac{1}{ {\bf l}_a^2 }    \right)
\ln \frac{{\bf l}_a^2}{ {\bf l}_b^2}
\right]
\notag \\
&
-
\left[
   3 + \left(  1 + \frac{n_f}{N_c^3} \right)
   \left( 
     1 - \frac{({\bf l}_a^2 + {\bf l}_b^2 )^2}{8{\bf l}_a^2  {\bf l}_b^2}
     - \frac{ (2{\bf l}_a^2 {\bf l}_b^2 -3 {\bf l}_a^4 - 3 {\bf l}_b^4  )}{16{\bf l}_a^4  {\bf l}_b^4} ({\bf l}_a \cdot {\bf l}_b )^2
   \right)
\right]
\notag \\
&
\times
\int_0^\infty dx \frac{1}{{\bf l}_a^2 + x^2 {\bf l}_b^2 } \ln \left|\frac{1 + x}{1 - x} \right|
\notag \\
&
+
\frac{ 2 ( {\bf l}_a^2 - {\bf l}_b^2)}{ ( {\bf l}_a - {\bf l}_b)^2 ( {\bf l}_a + {\bf l}_b)^2 } 
\left[
 \frac{1}{2} \ln \frac{ {\bf l}_a^2}{{\bf l}_b^2}
 \ln \frac{ {\bf l}_a^2 {\bf l}_b^2 ( {\bf l}_a - {\bf l}_b)^4 }{( {\bf l}_a^2 + {\bf l}_b^2)^4}
+
\left( \int_0^{-{\bf l}_a^2/{\bf l}_b^2} - \int_0^{-{\bf l}_b^2/{\bf l}_a^2} \right) dt
 \frac{\ln  (1 - t)}{t}
\right]
\notag \\
&
-
\left( 1 -  \frac{  ( {\bf l}_a^2 - {\bf l}_b^2)^2}{ ( {\bf l}_a - {\bf l}_b)^2 ( {\bf l}_a + {\bf l}_b)^2 }  \right)
\left[
 \left( \int_0^1 -  \int_1^\infty  \right) d z \frac{1}{({\bf l}_b - z {\bf l}_a )^2 } \ln \frac{( z {\bf l}_a)^2}{ {\bf l}_b^2}
\right]
\bigg \}
\end{align}
and
\begin{align}
  \label{eq:fkl_coll}
K_{\text{coll}} ({\bf l}_a, {\bf l}_b) =&
- \frac{1}{2} \int d^2 {\bf l} \, K^{\text{LO}}_{\text{BFKL}}  
( {\bf l}_a, {\bf l}) 
\, K^{\text{LO}}_{\text{BFKL}} ({\bf l}, {\bf l}_b) 
\, \ln{\frac{{\bf l}^2 }{ {\bf l}_b^2}}.
\end{align}

Following Ref.~\cite{Andersen:2003wy}, this representation of the
NLO BFKL kernel can be now used to solve iteratively the integral 
equation. The explicit solution for the gluon Green function then reads
\begin{align}
  \label{eq:iterative_greens}
 f(x,{\bf q}, {\bf k} )  =
x^{-\omega_\lambda (\bf q)}
&
 \bigg\{ \delta^{(2)} ( {\bf q} - {\bf k}) +
\sum_{n=1}^\infty \prod_{i = 1}^n \int d^2 {\bf l}_i 
\bigg[
\tilde{K}^{\text{real}}_\lambda ({\bf q} + \sum_j^{i-1} {\bf l}_j,{\bf q} + \sum_j^i {\bf l}_j  )
\notag \\
&
\times
\int_{ x_{i-1}}^1 \frac{d  x_i }{ x_i}
 x_i^{ -\omega_\lambda({ \bf q }+ \sum_{j =1}^i {\bf l}_j) 
+ \omega_\lambda({ \bf q }+ \sum_{j =1}^{i-1} {\bf l}_j ) } 
\bigg]
\delta^{(2)} ( {\bf q} + \sum_{j=1}^{n} {\bf l}_j -   {\bf k})\bigg\},
\end{align}
with $x_0 \equiv x$.  Note that this representation can be now implemented 
in a Monte Carlo event generator where all the information about each of the 
emitted particles is recorded. At NLO each iteration of the kernel, or 
each of the terms in the sum of Eq.~(\ref{eq:iterative_greens}), 
corresponds to one or two emissions well separated in rapidity from previous 
and subsequent clusters of particles. Inserting this function in the formula 
for the differential distributions will generate our exclusive observables. 

In the real emission kernel of Eq.~(\ref{eq:kernel_lamgda}), 
the two terms explicitly written 
in Eqs.~(\ref{eq:real_finite},\ref{eq:fkl_coll}) do not 
carry, apart from an overall $\bar{\alpha}_s^2 (\mu^2)$ factor, any 
renormalisation
scale dependence. This is different from the remaining part of the real
emission kernel and the gluon trajectory which contain the function 
$\xi$ of Eq.~(\ref{eq:omega_lambda}). This can be written as
\begin{equation}
  \label{eq:log_xi}
\xi(X ) ~=~ \bar{\alpha}_s(\mu^2) \left(   
    1 - \frac{\bar{\alpha}_s(\mu^2)}{4 N_c} \beta_0 \ln \frac{X}{\mu^2} + {\bar{\alpha}_s}(\mu^2) S \right),  \hspace{.4cm} \text{with} \hspace{.4cm}
S ~=~  \frac{1}{12}\left(4 - \pi^2 + 5 \frac{\beta_0}{N_c} \right).
\end{equation}
In this expression, the logarithmic term  can be absorbed into a
redefinition of the running of the coupling which corresponds to the
replacement of $\bar{\alpha}_s(\mu^2) $ by $\bar{\alpha}_s(X)$. 
The remaining, non-logarithmic term, can be identified as a common factor 
which appears when dealing with resummations of soft gluons~\cite{vierzehn}. 
The term $\bar{\alpha}_s \left( 1 + \bar{\alpha}_s S \right)$ is 
proportional to the two-loop cusp anomalous dimension. 
Generally, the appearance of this term offers the possibility to
change from the $\overline{\text{MS}}$ renormalisation scheme to the {\it
  Gluon-Bremsstrahlung} (GB) scheme. Such a change corresponds to a shift of
the Landau pole $\Lambda_{\text{GB}} = \Lambda_{\overline{\text{MS}}}
\exp{S \frac{2 N_c}{\beta_0}}$. Stability under this change of scheme 
offers a good tool to test our theoretical predictions. 

\section{Conclusions}
\label{sec:out}

In this letter we set the theoretical framework for a study of heavy
flavor production in central regions of rapidity in hadron-hadron 
collisions using $k_T$ factorisation at NLO.  While the heavy
flavor production vertex is kept at LO, the unintegrated gluon density 
is treated by taking into account the
full NLO corrections. The latter contain both the full NLO BFKL
evolution and the further corrections which arise due to the
asymmetric choice of energy scales, inherent to hadronic
cross-sections, and which can be understood as the onset of collinear
evolution from the soft hadrons to the hard production vertex.

The NLO BFKL Green function which, convoluted with the
(non-perturbative) proton impact factor, forms the NLO unintegrated 
gluon density has been presented in an iterative way which allows for a 
numerical evaluation using Monte-Carlo integration techniques. In future 
publications we will present results on this numerical 
implementation, together with fits of our unintegrated gluon density to 
deep inelastic data from HERA and predictions for heavy quark pair production 
at the Large Hadron Collider at CERN.

\subsubsection*{Acknowledgments}

M.H. thanks the ``Instituto de F{\'\i}sica Te\'orica UAM/CSIC'' at the 
Aut\'onoma University in Madrid and the Paul Scherrer Institut in Villigen 
for hospitality.

\end{document}